\documentclass[prd,twocolumn,superscriptaddress,floatfix,amsmath,amssymb,amsfonts,nofootinbib,longbibliography]{revtex4-2}

\usepackage{float} 

\usepackage[normalem]{ulem}
\usepackage[english]{babel}
\usepackage{graphicx}% Include figure files
\usepackage{dcolumn}% Align table columns on decimal point
\usepackage{faktor}
\usepackage{bm}% bold math
\usepackage{blindtext}
\usepackage{verbatim}
\usepackage{relsize}
\usepackage{mathrsfs}
\usepackage{musicography}
\usepackage{amsmath}
\usepackage{blindtext}
\usepackage{cancel}
\usepackage{physics}
\usepackage{epstopdf}
\usepackage{mathtools}
\usepackage{blindtext}
\usepackage{tensor}
\usepackage{color}
\usepackage[usenames,dvipsnames]{pstricks}
\usepackage{epsfig}
\usepackage{pst-grad} % For gradients
\usepackage{pst-plot} % For axes
\usepackage{hyperref}
\usepackage{verbatim}
\usepackage{slashed}
\usepackage{dsfont}

\newcommand{\mf}{\mathsf}%\mathsf is too long

\newcommand{\ii}{\mathrm{i}}

\newcommand{\tc}[1]{\textsc{#1}}

\newcommand{\tbb}[1]{\textcolor{black}{#1}}

\allowdisplaybreaks[1] %<<<<<<<<<<<<<<<<<<<<<<
\begin{document}

\title{What gravity mediated entanglement can really tell us about quantum gravity}

\author{Eduardo Mart\'in-Mart\'inez}
\email{emartinmartinez@uwaterloo.ca}

\affiliation{Department of Applied Mathematics, University of Waterloo, Waterloo, Ontario, N2L 3G1, Canada}
\affiliation{Perimeter Institute for Theoretical Physics, Waterloo, Ontario, N2L 2Y5, Canada}
\affiliation{Institute for Quantum Computing, University of Waterloo, Waterloo, Ontario, N2L 3G1, Canada}

\author{T. Rick Perche}
\email{trickperche@perimeterinstitute.ca}

\affiliation{Department of Applied Mathematics, University of Waterloo, Waterloo, Ontario, N2L 3G1, Canada}
\affiliation{Perimeter Institute for Theoretical Physics, Waterloo, Ontario, N2L 2Y5, Canada}
\affiliation{Institute for Quantum Computing, University of Waterloo, Waterloo, Ontario, N2L 3G1, Canada}

\begin{abstract}
    %We revisit the Bose-Marletto-Vedral (BMV) table-top experiment for quantum gravity using particles in superposition of trajectories locally coupled to a classical gravitational field. We describe a local interaction by taking into account the propagation of the weak gravitational field generated by each particle, and show that a classical gravitational field can be responsible for the entanglement. We then compare this scenario with the case where the gravitational field is treated as a quantum field, where we also obtain a vacuum contribution to the entanglement. Comparing the two cases, we conclude that unless the entanglement contribution from the gravitational field can be measured, one cannot claim quantum nature of the gravitational field based on the results of experiments. We also precisely quantify the regimes where the major entanglement source from the particles can be attributed to the quantum nature of the gravitational field, namely, when the masses are mostly spacelike separated throughout their interaction. This allow us to adapt the experiment in order to claim true quantum nature of the gravitational field.
    We revisit the Bose-Marletto-Vedral (BMV) table-top  experimental proposal---which aims to witness quantum gravity using gravity mediated entanglement---analyzing the role of locality in the experiment. We \tbb{first} carry out a fully quantum modelling of the interaction of matter and gravity and \tbb{then show in what way gravity mediated entanglement in the BMV experiment could be accounted for without appealing to quantum degrees of freedom of the gravitational field.} \tbb{We discuss what assumptions are needed in order to interpret the current BMV experiment proposals as a proof of quantum gravity, and also} identify the modifications that a BMV-like experiment \tbb{could} have in order \tbb{ to serve as proof of quantum gravity without} \tbb{having to assume the existence of a local mediators in the gravitational field}.
\end{abstract}

\maketitle

%\section*{}

%\textbf{\textit{{Introduction.- }}} 

\textbf{\textit{Introduction.--}} A full formulation of quantum gravity is among the most coveted theories in physics. However, we  lack experimental testbeds to probe gravity in the quantum regime. Recently, exciting proposals for tabletop experiments combining quantum mechanics and gravity have been put forward by Bose et. al~\cite{B} and Marletto and Vedral~\cite{MV} (the so-called BMV experiment). These proposals have captured the eye of the  community for their potential to provide \tbb{insights on} the \tbb{interaction of gravity and quantum matter}. 

The experiment consists of two particles, each prepared in a superposition of two different trajectories. The two particles interact only  gravitationally. The rough idea is that the different superposed paths  generate different gravitational fields, which can entangle the particles. Despite the inherent experimental challenges to isolate the particles  while shielding them from decoherence, we are close to having technology that will allow the experiment in the near future~\cite{B,MV,marios,marios2,enhance2,enhance1,markus1,markus2}. Although it is exciting to have an experiment that can explore aspects of  gravity when quantum  theory is relevant,  some possible objections to the fact that the BMV experiment can reveal the quantum nature of gravity  exist in the literature~\cite{CharisHu,treta,tretaReply,tretaRereply,Anastopoulos2021,Julen}. This indicates that  the conclusions that can be drawn from the experiment require careful analysis. 
%The goal of the BMV experiment is to verify that the masses will be entangled after the gravitational interaction. 
For example, it has been claimed that if the gravitational field is capable of entangling the masses, %then the gravitational field must be quantum. 
then either we must abandon the principle of locality, or the gravitational field must be quantum~\cite{MVwhen,BMVqft}. In this paper we present a series of arguments \tbb{which qualify this statement}. \tbb{We} discuss that while the experiment, as proposed, can prove that gravity can establish a quantum channel between the particles, it cannot decide whether gravity has quantum degrees of freedom. Furthermore, we propose a modification to the experiment so that it can %\sr{addresses the}
demonstrate the fundamental quantum behaviour of gravity.

\tbb{It has been argued} that \tbb{when one tries to describe the} coupling of classical and quantum systems\tbb{,} \tbb{one is faced with \emph{theoretical} inconsistencies} (see, e.g., ~\cite{Terno2006,raul}). \tbb{From this perspective,} theories where matter is quantum and gravity is classical \tbb{cannot be} fundamental. However, the theoretical argument alone cannot be used to claim that gravity is quantum%(e.g., the interaction between matter and gravity might perhaps be described by a different framework outside quantum theory)
. Instead, for an experimental claim of quantumness, one requires observing specific markers of quantum behaviour. While different authors may disagree on what conditions are sufficient to prove that a system is quantum, there are some uncontroversial properties of a system that, if observed, would prove its quantumness. For example, observing Wigner negativity~\cite{wignerNeg,PRXwigner}, or violations of Bell inequalities~\cite{PRXwigner,bellInVio} \tbb{in the system itself}.

In this light, our goal is twofold. First, we will discuss that \tbb{while} the \emph{current} proposals of the BMV experiment \tbb{can explore whether gravity can mediate entanglement between two quantum systems, it cannot directly witness quantum degrees of freedom of gravity, unless one also assumes the existence of local gravitational mediators.} That is, we will argue that the mere observation that two masses (in a quantum superposition of trajectories) get entangled through the gravitational interaction is not sufficient to prove the existence of gravitational quantum degrees of freedom. Second, we will also identify under which regimes a BMV-like experiment could reveal \tbb{more about the} quantum \tbb{behaviour} of gravity \tbb{with fewer assumptions}.

\textbf{\textit{A fully quantum description of the experiment.--}} We first employ a quantum field theoretic description of gravity in the BMV experiment, where the  masses are coupled to linearized quantum gravity: a weak field limit of the gravitational field which can be quantized~\cite{DeWittQG,verch2022}. This weak-field limit description should be a prediction of any theory of quantum gravity. Consider the metric
\begin{equation}\label{eq:metricExp}
    g_{\mu\nu} = \eta_{\mu\nu} + {\sqrt{16\pi G}}\,h_{\mu\nu},
\end{equation}
where $\eta_{\mu\nu} = \text{diag}(-1,1,1,1)$ and $h_{\mu\nu}$ is a metric perturbation with units of energy. We choose these conventions so that the field propagators do not pick up factors of $G$. We quantize the gravitational perturbations as the quantum field $\hat{h}_{\mu\nu}(\mf x)$. The predictions of the theory are entirely determined by the field's Wightman and Feynman propagator distributions,
\begin{align}
    W_{\mu\nu\alpha'\beta'}(\mf x,\mf x') &= \langle \hat{h}_{\mu\nu}(\mf x) \hat{h}_{\alpha'\beta'}(\mf x')\rangle \nonumber\\&= \frac{\ii}{2}E_{\mu\nu\alpha'\beta'}(\mf x,\mf x') + \frac{1}{2}H_{\mu\nu\alpha'\beta'}(\mf x,\mf x'),\nonumber\\
    G_{\mu\nu\alpha'\beta'}(\mf x,\mf x') &= \langle \mathcal{T} \hat{h}_{\mu\nu}(\mf x) \hat{h}_{\alpha'\beta'}(\mf x')\rangle\nonumber \\*&={-}\frac{\ii}{2}\Delta_{\mu\nu\alpha'\beta'}(\mf x,\mf x') + \frac{1}{2}H_{\mu\nu\alpha'\beta'}(\mf x,\mf x'),\label{eq:G}
\end{align}
where $\mathcal{T}$ denotes the time ordering operation, $E_{\mu\nu\alpha'\beta'}(\mf x,\mf x')$ is the causal propagator (the advanced minus retarded Green's functions), $\Delta_{\mu\nu\alpha'\beta'}(\mf x, \mf x')$ is the radiation Green's function (the retarded \emph{plus} advanced Green's function), and $H_{\mu\nu\alpha'\beta'}(\mf x,\mf x')$ is the Hadamard distribution. It is important to distinguish which of these functions are a consequence of the field quantization, and which ones are also present for a classical field. The (state independent) propagator $E_{\mu\nu\alpha'\beta'}(\mf x,\mf x')$ is proportional to the field's commutator, and as such, it is genuinely quantum~\cite{quantClass}. On the other hand, $ \Delta_{\mu\nu\alpha'\beta'}(\mf x,\mf x')$ is associated with the \emph{classical} propagation of perturbations between the events $\mf x$ and $\mf x'$. $H_{\mu\nu\alpha'\beta'}(\mf x,\mf x')$ contains all the state dependent part of  $\text{both }W_{\mu\nu\alpha'\beta'}(\mf x,\mf x')\text{ and }G_{\mu\nu\alpha'\beta'}(\mf x,\mf x')$, including the fundamentally quantum contributions from the vacuum and the presence of entanglement in the field (see~\cite{quantClass}).  

\begin{figure}[h!]
    \centering
    \includegraphics[width=8.6cm]{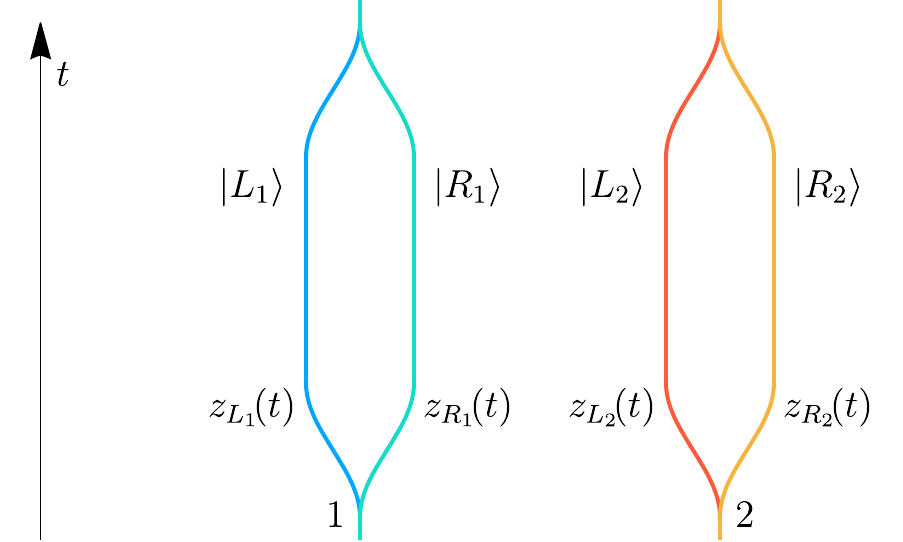}
    \caption{Schematic representation of the BMV setup, where two particles labelled by $i=1,2$ can undergo a superposition of two trajectories, $\mf z_{L_i}(t)$ and $\mf z_{R_i}(t)$, corresponding to quantum states $\ket{L_i}$ and $\ket{R_i}$.}
    \label{fig:BMV}
\end{figure}

To describe the BMV experiment, consider two pointlike particles labelled by $i\in\{1,2\}$ with masses $m_1$ and $m_2$, whose centres of mass are quantum and can undergo two possible trajectories each, $\mf z_{R_i}(t)$ and $\mf z_{L_i}(t)$. We associate each possible trajectory to states $\ket{R_i}$ and $\ket{L_i}$ (see Fig.~\ref{fig:BMV}). The quantum description of the experiment involves two interactions of each particle with the quantum gravitational field according to the interaction Hamiltonian density
\begin{equation}
    \hat{\mathcal{H}}_I(\mf x) = -{\sqrt{4\pi G}}\!\!\!\!\!\sum_{p_i\in\{L_i,R_i\}}\!\!\!\!\! \ket{p_i}\!\!\bra{p_i}T_{p_i}^{\mu\nu}(\mf x)\hat{h}_{\mu\nu}(\mf x),
\end{equation}
where
\begin{equation}\label{eq:Tpi}
    T_{p_i}^{\mu\nu}(\mf x) = m_i\, u_{p_i}^\mu(t) u_{p_i}^\nu(t) \frac{\delta^{(3)}(\bm x - \bm z_{p_i}(t))}{u_{p_i}^0(t)\sqrt{-g}}
\end{equation}
is the stress-energy tensor corresponding to each path $\mf{z}_{p_i}(t)$ for the particles with four-velocities $u_{p_{{i}}}^\mu(t)$.

% Similar treatments have been used to describe particle detectors undergoing superposition of trajectories (see e.g.~\cite{achimDelocalizedHarvesting,foo}).
The interaction Hamiltonian is given by   \mbox{$\hat{H}_I(t) =\int\dd^3\bm x\,\hat{\mathcal{H}}_I(\mf x)=$}
\begin{equation}\label{eq:Hq}
    -{\sqrt{4\pi G}}\!\!\!\!\!\sum_{{\substack{p_1\in\{L_1,R_1\}\\p_2\in\{L_2,R_2\}}}}\!\!\!\!\! \ket{p_i}\!\!\bra{p_i}m_i\frac{u_{p_i}^{\mu}(t)u_{p_i}^{\nu}(t)}{u_{p_i}^0(t)}\hat{h}_{\mu\nu}(\mf z_{p_i}(t)),
\end{equation}
and the unitary time evolution operator is then given by $\hat{U}_I = \mathcal{T}\exp(-\ii \int \dd t \,\hat{H}(t)),$ where $\mathcal{T}\exp$ denotes the time-ordered exponential. We analyze the experiment using perturbation theory, and  %\sr{The perturbative calculations of the time evolution operator} 
the calculations are standard (see Appendix~\ref{app:perturbation}).

%\sr{In order} 
The specific BMV setup considers both particles to be in a superposition of the two paths, so that the initial state of the system is
\begin{equation}\label{eq:psi0}
    \ket{\psi_0} = \frac{1}{\sqrt{2}}\left(\ket{L_1}+\ket{R_1}\right)\otimes\frac{1}{\sqrt{2}}\left(\ket{L_2}+\ket{R_2}\right),
\end{equation}
corresponding to the density operator
\begin{equation}
    \hat{\rho}_0 = \ket{\psi_0}\!\!\bra{\psi_0} = \frac{1}{4}\sum_{{\substack{p_1\in\{L_1,R_1\}\\p_2\in\{L_2,R_2\}}}} \ket{p_1p_2}\!\!\bra{q_1q_2}.
\end{equation} 
To obtain the final state of the  particles we assume that the field is in the vacuum state. After the interaction, one must trace-out the  gravitational degrees of freedom. This results in a mixed state for the two masses, because the particles become entangled with the gravitational field \tbb{itself}. To simplify the result, we assume that all trajectories are related by rotations and translations in space, so that the local vacuum effect in each trajectory is the same. Under this assumption, the 
leading order negativity is computed in Appendix~\ref{app:perturbation}
\begin{align}\label{eq:Nq}
    \mathcal{N}_{\textsc{g}} \!=\! {\pi G}\Big(\Big|G_{L_1L_2}\!\!+\!G_{R_1R_2}\!\!-\!G_{L_1R_2}\!\!-\!G_{R_1L_2}\Big|\!-\!\mathcal{L}\Big) +\mathcal{O}(G^2),
\end{align}
where $\mathcal{L}$ is a local noise term associated to the particle's interaction with the gravitational field vacuum---which are given by integrals of $W_{\mu\nu\alpha'\beta'}(\mf x, \mf x')$---and 
\begin{equation}
    G_{p_1p_2} = \int \dd V\dd V' T_{p_1}^{\mu\nu}(\mf x) G_{\mu\nu\alpha'\beta'}(\mf x,\mf x') T_{p_2}^{\alpha'\beta'}(\mf x'),
\end{equation}
where $\dd V$ denotes the invariant spacetime volume element. As per the discussion after Eq. \eqref{eq:G}, the role played by the quantum degrees of freedom of the gravitational field in the entanglement mediation is then encoded in the dependence of $\mathcal{N}_\tc{G}$ on real part of $G_{\mu\nu\alpha'\beta'}(\mf x, \mf x')$ and the local noise terms $\mathcal{L}$.

\textbf{\textit{Quantum-controlled classical fields can entangle.--}} We now present a description of the BMV experiment which does not rely on any quantum property of gravity and is compatible with relativity. \tbb{We} will show that this description is able to precisely predict the expected results of the experiment in the regimes previously proposed~\cite{B}. In this model, we consider linearized perturbations as in Eq.~\eqref{eq:metricExp}, and write the classical solution of the linearized Einstein's equations for a source $T_{\mu\nu}$ as
\begin{equation}\label{eq:prop}
    h^{\mu\nu}(\mf x) = {\sqrt{4\pi G}}\int \dd V' G_R^{\mu\nu}{}_{\alpha'\beta'}(\mf x,\mf x') T^{\alpha' \beta'}(\mf x'),
\end{equation}
where $G_R^{\mu\nu}{}_{\alpha'\beta'}(\mf x,\mf x')$ denotes a retarded Green's function for the linearized Einstein's equations. Each possible path combination $\ket{R_1R_2}, \ket{R_1L_2}, \ket{L_1R_2}$, $\ket{L_1L_2}$ will be associated with a different classical gravitational interaction between the masses. In this sense, this description is a quantum-controlled classical model for the gravitational field, where gravity has no quantum degrees of freedom. The interaction Hamiltonian can be written as
\begin{equation}\label{eq:Hclass}
    \hat{H}_I(t) = \sum_{{\substack{p_1\in\{L_1,R_1\}\\p_2\in\{L_2,R_2\}}}} {\Phi_{p_1p_2}(t)}\ket{p_1p_2}\!\!\bra{p_1p_2} ,
\end{equation}
where $\Phi_{p_1p_2}(t)$ denotes the retarded propagation of the classical gravitational field between the two particles undergoing trajectories $\mf z_{p_1}(t)$ and $\mf z_{p_2}(t)$, i.e.,
\begin{align}
    {\Phi_{p_1p_2}(t) = -{\sqrt{{\pi G}}}\!\displaystyle{\int} \dd^3\bm x \! \left(T^{\mu\nu}_{p_1}(\mf x) h^{p_2}_{\mu\nu}(\mf x)+T^{\mu\nu}_{p_2}(\mf x) h^{p_1}_{\mu\nu}(\mf x)\right)\!,}
\end{align}
{where}
\begin{align}
    {h_{p_i}^{\mu\nu}(\mf x) = {\sqrt{4\pi G}}\int \dd V' G_R^{\mu\nu}{}_{\alpha'\beta'}(\mf x,\mf x') T^{\alpha'\beta'}_{p_i}(\mf x')}
\end{align}
{denotes the field generated by particle $i$ undergoing path $p_i$ and }
\begin{equation}\label{eq:Tpi}
    T_{p_i}^{\mu\nu}(\mf x) = m_i\, u_{p_i}^\mu(t) u_{p_i}^\nu(t) \frac{\delta^{(3)}(\bm x - \bm z_{p_i}(t))}{u_{p_i}^0(t)\sqrt{-g}}.
\end{equation}
This approach is the relativistic (and thus causal) unapproximated version of the interaction \mbox{$Gm_1m_2/|\hat{\bm x}_1 - \hat{\bm x}_2|$}, which \emph{does not rely on any quantum property of the gravitational field}. \tbb{Even though the quantum-controlled gravitational field does not have definite values at every point of spacetime, we believe that this description} does not involve \tbb{a quantum} superposition of gravitational fields \tbb{in the traditional sense,} since there is no Hilbert space for gravity: a quantum-controlled classical field is certainly not itself a quantum field. This is in tension with the wording used in e.g.,~\cite{marios,relBMV,CaslavRecommendation}. %Notice that this is \emph{not} a superposition of gravitational fields, as there is no Hilbert space for the state of the gravitational field: a quantum-controlled classical field is not a quantum field.

{The interaction~\eqref{eq:Hclass} defines a quantum channel between the particles that respects relativity. However, establishing a quantum channel cannot be taken as proof that gravity has quantum degrees of freedom: it is merely a consequence of having quantum sources and not of any assumptions about the field (see~\cite{Perro}). Indeed, this model does not assume that the gravitational field has degrees of freedom of its own that carry information between the particles (i.e. gravity is not an active mediator). We  use the term quantum-controlled classical to refer to~\eqref{eq:Hclass} because it associates to each state of the particles the classical field sourced by each particle undergoing each path. %The name ``classical'' is given in contrast to the model presented later where the field has local quantum degrees of freedom and acts as a mediator for the interaction. }

Since the gravitational interaction is implemented by a classical field, $\hat{H}_I(t)$ commutes with itself at different times. The %\sr{time evolution}
time-evolution operator %\sr{will be given by} 
is
\begin{align}
    \hat{U}_I &= \exp({-\ii \int\dd t\,\hat{H}_I(t)})\nonumber\\
    &=  \sum_{{\substack{p_1\in\{L_1,R_1\}\\p_2\in\{L_2,R_2\}}}} e^{{2\pi\ii G} \Delta_{p_1p_2}}\ket{p_1 p_2}\!\!\bra{p_1p_2},
\end{align}
where $\Delta_{p_1p_2}$ is a double integral in spacetime of the retarded plus advanced propagator contracted with the stress-energy tensor of the sources corresponding to each path:
\begin{equation}
    \Delta_{p_1p_2} \coloneqq \displaystyle{\int} \dd V \dd V' T_{p_1}^{\mu\nu}(\mf x)\Delta_{\mu\nu}{}_{\alpha'\beta'}(\mf x,\mf x') T_{p_2}^{\alpha'\beta'}(\mf x').
\end{equation}

%In fact the quantum notion of locality has to be violated anytime that entanglement between two systems is generated. For example, a local interaction at a single point in spacetime between two quantum fields that does not commute with their free dynamics generates non-local (in the quantum sense) unitary evolution even though the interaction is local in spacetime. 

Using the initial state for the particles given in Eq.~\eqref{eq:psi0}, we obtain the following final density operator after the interaction
\begin{equation}
    \hat{\rho}_{c} = %\hat{U}_I\hat{\rho}_0\hat{U}_I^\dagger =
    \frac{1}{4}\sum_{{\substack{p_1\in\{L_1,R_1\}\\p_2\in\{L_2,R_2\}}}} \!\!\!e^{ {2\pi \ii G}( \Delta_{p_1p_2}-\Delta_{q_1q_2})}\ket{p_1p_2}\!\!\bra{q_1q_2}.
\end{equation}
The entanglement between the two particles can be evaluated through the negativity~\cite{footNeg} of the state $\hat{\rho}_c$, which reads
\begin{align}
    \mathcal{N}_{\textsc{c}} = \frac{1}{2}\sin({\pi G} \Big|{\Delta_{L_1L_2}\!\!+\!\Delta_{R_1R_2} \!\!-\! \Delta_{L_1R_2} \!\!-\! \Delta_{R_1L_2}}\Big|)\nonumber\\
    = {\frac{\pi G} {2}}\Big|{\Delta_{L_1L_2}\!\!+\!\!\Delta_{R_1R_2}\!\! -\!\! \Delta_{L_1R_2}\!\! -\!\! \Delta_{R_1L_2}}\Big|+\mathcal{O}(G^2).\label{eq:Nc}
\end{align}
With the typical choice of paths for the BMV experiment, the above quantity is non-zero. Therefore, it is possible to model entanglement creation \tbb{with a relativistic interaction that does not involve local quantum degrees of freedom of the gravitational field}.  Moreover, Eq.~\eqref{eq:Nc} is exactly the result obtained when one sets the fundamentally quantum parts of the two-point functions (i.e., $E_{\mu\nu\alpha'\beta'}(\mf x,\mf x') \text{ and } H_{\mu\nu\alpha'\beta'}(\mf x, \mf x')$) to $0$ in Eq.~\eqref{eq:Nq}. That is, this model is the limit of the quantum field theoretical description when one gets rid of the quantum degrees of freedom of the gravitational perturbations. Overall, this causal description does not involve \tbb{local quantum degrees of freedom for} gravity and yet it is able to predict the expected gravity-mediated entanglement generation in the BMV experiment.

\textbf{\textit{Comparing the models.--}} Comparing Eqs.~\eqref{eq:G} and \eqref{eq:Nq} with \eqref{eq:Nc}, we see that, apart from the vacuum noise $\mathcal{L}$ that appears in the quantum case, the entanglement acquired in the quantum field description is larger than the one obtained in the quantum-controlled classical description. {In the regime of long interaction times, the noise term $\mathcal{L}$ is insignificant compared to the effect of the propagators, so that it can be neglected~\cite{quantClass}.} The contribution to the negativity due to the imaginary part of the Feynman propagator can be associated with entanglement mediated by communication via the field~\cite{ericksonNew,quantClass}, while the real part of the propagator is associated with the entanglement \emph{extracted} from the vacuum state of the gravitational field (in relativistic quantum information this phenomenon is known as entanglement harvesting~\cite{Pozas-Kerstjens:2015,ericksonNew,remi}). In this sense, the real part of the propagator ($H_{\mu\nu\alpha'\beta'}(\mf x, \mf x')$) is the one that captures the entanglement associated {to} the quantum degrees of freedom of the gravitational field. Violating Bell inequalities with this entanglement would then be proof of the quantum behaviour of gravity. In \tbb{most} setups where the paths are causally connected, the contribution of the real part of the propagator is negligible compared to its imaginary counterpart ($\Delta_{\mu\nu\alpha'\beta'}(\mf x, \mf x')$). In particular, the proposed implementations of the BMV experiment use masses with smallest separation of the order of $L\sim10^{-6}\text{m}$ and interaction times of the order of $T\sim1\text{s}$. For these parameters we find that the imaginary part contribution of the propagator is {$10^{14}$} times larger than its real part, \tbb{making the BMV experiment agnostic to the existence of local quantum degrees of freedom for the gravitational field}. 

In order to discuss the interpretation of the experiment, it is also important to distinguish between two fundamentally different notions of locality. The first notion comes from the description of spacetime %\sr{(e.g. general relativity)} 
and is deeply linked with causality. It states operations happen at events in spacetime, and do not affect other events which are causally disconnected from them. We will call this notion \textit{event locality}. The second notion of locality comes from quantum mechanics, and states that operations that independently affect two quantum systems must be separable. We call this notion \textit{system locality}~\cite{ThomasFlaminiaAndJohn}. The notion of system locality alone is agnostic about causal structure or any underlying notion of spacetime. Although these notions of locality are different there are particular frameworks in which we link the two. E.g., in quantum field theory (QFT) the postulate of microcausality prevents operations in local systems from violating the notion of event locality prescribed by relativity. % In QFT it is assumed that the notion of system locality is complemented with the inability of system-local operations to influence events that are not casually connected.

The distinction between the two notions of locality is particularly important when talking about local operations and classical communication (LOCC). While the notion of system locality is operationally captured by the `L' and the `O', we often define what we mean by classical communication (the `CC') in terms of event locality. After all, what is classical communication if not sending information from one event to another?

%The different frameworks in which we link event locality and system locality are defined by the way in which communication between events can happen. For example, in non-relativistic quantum mechanics we would have the Galilean notion of event locality. If we impose that the notion of event locality is given by relativity, the framework that we build is QFT. And in general, for any GPT (generalized probabilistic theory) we would have to equip the system locality of quantum mechanics with a notion of event locality.

Many works in the literature which argue that the BMV experiment can be used to probe the quantum nature of gravity  use an argument based on LOCC. In essence, the argument goes as follows: LOCC does not increase the entanglement between quantum systems, thus, if the masses interact only gravitationally and get entangled, the gravitational field which mediates the interaction is going beyond `CC'~\cite{ansToVedral}, hence the field cannot be classical~\cite{B,MV,MVwhen,BMVqft}. However, these arguments assume a relationship between event and system locality in order to reach their conclusion: they assume that, in order to get the masses entangled, a mediator is required in order not to violate system locality, ruling out a direct interaction between the masses. This assumption can be reworded as follows:  \textit{mass A couples to the field and then the field carries quantum information to mass B, or  otherwise we would have action-at-a-distance}. However, the assumption that the gravitational interaction is implemented through system-local operations is not based on first principles. It is reasonable to demand that gravity must satisfy event locality to prevent action-at-a-distance, but it is not unthinkable to consider that it may not necessarily %\sr{satisfy system locality} 
be \mbox{system-local} since this notion is operational rather than fundamental. 

We know that %\sr{a framework like QFT for the description of} 
if we use QFT to describe gravity,  the %\sr{gravitational} 
interaction will be both \mbox{event-local} and system-local. %However there are other possible models for gravity that are not system-local but still, they are event-localthus assuming that the two notions are connected from the getgo cannot be used to prove that the theory is quantum. 
However, assuming %\sr{to assume}
this relationship between the two notions of locality {to interpret the results of the experiment} %\sr{from the beginning} 
implicitly assumes \textit{a-pirori} that the system is described by a framework like QFT. This is not satisfactory if our objective is to prove the quantum nature of a relativistic theory. For instance, the fact that a classical Coulomb potential can entangle two charges 
is not why we believe that the electromagnetic field is quantum{: $\faktor{q}{\hat{r}}$ is clearly not a quantum field.} The electromagnetic field has only been proved to be quantum when QFT was \emph{required} to model experiments which no classical theory could account for, not when the quantum description for the hydrogen atom was verified.

Importantly, the time evolution generated by the quantum-controlled classical field interaction is not local in the operational quantum sense (that is, $\hat{U}_I \neq \hat{U}_1\otimes \hat{U}_2$ is not \mbox{system-local}), although the interaction is intrinsically local in the relativistic sense and thus satisfies event locality. 
This means that time evolution implemented by $\hat{U}_I$ can create entanglement, even though the gravitational field \tbb{does not have quantum degrees of freedom} and the interaction is \mbox{event-local} in this description. The notion of event locality is the one that comes from first principles and captures no-action-at-a-distance: the field only interacts with the particles locally at each instant of time, due to the causal retarded propagation. \tbb{This means that seeing the BMV experiment establish a quantum channel between the masses and assuming Lorentz invariance is not enough to infer the existence of local quantum degrees of freedom of gravity: one would also need to assume that the interaction is implemented by local mediators.}

\textbf{\textit{A proposal for witnessing quantum gravity with the BMV experiment--}} Overall, the BMV experiment {can} be described \tbb{either by treating gravity as a quantum field, or by considering it to be a quantum-controlled field devoid of quantum degrees of freedom}. Although these descriptions yield different predictions (therefore the experiment \textit{does} have the potential to acknowledge quantum gravitational \tbb{degrees of freedom}), they can only be experimentally distinguished in specific regimes.

It is then possible to adapt the experiment to \tbb{directly witness} the quantum \tbb{degrees of freedom} of \tbb{the} gravitational field. If  it could be implemented for times of the order of the light-crossing time of the separation between paths, then the \tbb{Hadamard term} of the Feynman propagator would be larger than its imaginary part, and entanglement present in the gravitational field would be meaningful \tbb{in} the experiment. \tbb{If} the experiment can be adapted to implement these short interaction times (or can be sensitive  to changes in negativity of the order of $10^{-14}$ with the current proposed parameters~\cite{B}) one can claim to witness \tbb{quantum degrees of freedom} of the gravitational field \tbb{with no extra assumptions}. 

On the other hand, it would be unfair to say that the BMV experiment---even in the regimes considered in~\cite{B,MV}---gives no information about the interplay of gravity and quantum mechanics. If the gravitational field proves to be able to entangle the particles, we experimentally confirm that {gravity can mediate quantum channels between masses.} {In fact, if the BMV experiment confirms that gravity can mediate entanglement, it would rule out many classical descriptions for gravity coupled to quantum matter (such as some of the examples in~\cite{7nonstandard}).} {However, as we discussed, this is different from proving that gravity itself has quantum degrees of freedom.}

%, as we discussed.% such as stochastical gravity or passive quantum gravity~\cite{PQG95,PassiveQG2003,PQG2004,Stochastic2008}

%It also has the potential to rule out some particular non-quantum models of gravity (see~\cite{7nonstandard}).

%In fact, if the BMV experiment confirms that gravity can mediate entanglement, it would rule out many classical descriptions for gravity (see~\cite{7nonstandard}).

\textbf{\textit{Conclusions--}} We discussed the implications of the BMV experiment for quantum gravity%\sr{ in different regimes}
. If the experiment is implemented for long times, classical gravity {mediates} entanglement between the masses. This can give us information about how gravity couple to quantum systems, but {may} not be argued as proof of \tbb{existence of local quantum degrees of freedom} of the gravitational field. If on the other hand the experiment can be performed %\tb{as fast as}
in times of the order of the light-crossing time between the paths, then the entanglement between the particles would be significantly affected by the entanglement previously present in the gravitational field. This could be used to experimentally assess whether gravity admits a quantum field theoretic description or not. 

In order to adapt the previously proposed experiment so that the entanglement of the masses is due to the quantum nature of the field, either the spatial separation between the trajectories would have to be increased, or the interaction time would have to be decreased to guarantee spacelike separation. %\sr{Another way of witnessing quantum gravitational behaviour would be to have} 
Alternatively, one could also see quantum behaviour with  enough sensitivity in the experiment to distinguish between the  \tbb{quantum-controlled} (Eq.~\eqref{eq:Nc}) and the quantum \tbb{field result} (Eq.~\eqref{eq:Nq}).  %\sr{In fact,} 
It is even possible to find regimes where the quantum prediction \tbb{yields} less entanglement than the classical model predicts. This is due to the fact that for short interaction times, the local vacuum noise term $\mathcal{L}$ can dominate the sum in Eq.~\eqref{eq:Nq}. An effect that is well known within entanglement harvesting~\cite{Reznik2003,Pozas-Kerstjens:2015,ericksonNew,remi}.

While the physical relevance of the  %\sr{proposal} 
{BMV experiment} is unquestionable, we argued that the experiment as prescribed is not enough to assert \tbb{(without further assumptions)} the existence of local quantum \tbb{degrees of freedom in}  the gravitational field, unless taken to regimes where a \tbb{relativistically} local \tbb{quantum-controlled} model  cannot predict the entanglement acquired by the masses.  %We have here proposed a table-top experiment which can indeed probe the quantum nature of the gravitational field.
While we do not want to subtract from the many merits of the BMV proposal, it is important to qualify the range of implications of its experimental realization \tbb{and the assumptions needed in order to consider it a full experimental proof of quantum gravity. While the assumption that gravity is mediated by some kind of `graviton' degree of freedom may seem easy to accept for most, we believe it is important to emphasize that BMV experiment (in the regimes that it was originally proposed) cannot directly witness these `gravitons'. We also discussed that it is not enough to assume relativistic causality/Lorentz invariance in order to bypass the assumption of local mediators}. \tbb{We have shown, however, that the experiment can be adapted to not to have to rely on this assumption, although at the price of making it more experimentally challenging.}

%Finally, it is important to mention that it is possible for the BMV experiment to be adapted so that true quantum effects of the gravitational field can be accessed. It is enough that the quantum contributions (the real part of the Feynman propagators) to the negativity in Eq. \eqref{eq:Nq} are similar or larger than the classical ones (associated to the imaginary part of the propagators). This can be achieved if, for instance, the trajectories are distinct only in approximately spacelike separated regions, where the particles would extract entanglement from the vacuum of the gravitational field, rather than using it to communicate. This protocol is well known in the literature, and has been explored in many setups with different fields. In fact, in~\cite{remi}, it is proposed that entanglement harvesting from the gravitational field could be used to witness quantum behaviour of gravity if spacelike separated systems can get entangled. A precise quantification of the entanglement that can be acquired from the gravitational field using spacelike separated atoms was studied in detail in. Overall, one can only confidently claim that the gravitational field possesses quantum degrees of freedom if it is possible to \emph{extract} entanglement from it, but not if it can be used to entangle two quantum systems.

\vspace{5mm}

\begin{acknowledgments}

The authors are very grateful to Markus Aspelmeyer, \v{C}aslav Brukner\tbb{,} Flaminia Giacomini\tbb{, Sougato Bose, and Carlo Rovelli} for our very helpful extensive discussions on the BMV experiment and its interpretations.
The authors thank Julen Simon Pedernales, Martin B. Plenio and Kirill Streltsov for the insightful and helpful discussions about quantum channels and interactions. The authors also acknowledge Daniel Grimmer, Jos\'e Polo-G\'omez, Adam Teixid\'o-Bonfill, Bruno de S. L. Torres, Kelly Wurtz  for valuable discussions and feedback. T. R. P. acknowledges support from the Natural Sciences and Engineering Research Council
of Canada (NSERC) via the Vanier Canada Graduate Scholarship. E. M-M. is funded by the NSERC Discovery program as well as his Ontario Early Researcher Award. Research at Perimeter Institute is supported in part by the Government of Canada through the Department of Innovation, Science and Industry Canada and by the Province of Ontario through the Ministry of Colleges and Universities. Perimeter Institute and the University of Waterloo are situated on the Haldimand Tract, land that was promised to the Haudenosaunee of the Six Nations of the Grand River, and is within the territory of the Neutral, Anishnawbe, and Haudenosaunee peoples.

\end{acknowledgments}

%\begin{comment}
\appendix

\onecolumngrid

\section{The retarded propagator of the gravitational field}

The retarded propagator $G_R^{\mu\nu}{}_{\alpha'\beta'}(\mf x,\mf x')$ can be written as
\begin{equation}
    G_R^{\mu\nu}{}_{\alpha'\beta'}(\mf x,\mf x') = \frac{1}{2\pi}\theta(t-t') \delta\left((t-t')^2 - |\bm x - \bm x'|^2\right)\mathcal{P}^{\mu\nu}{}_{\alpha'\beta'} = \frac{1}{4\pi | \bm x - \bm x'|} \delta(t-t' - |\bm x  - \bm x'|)\mathcal{P}^{\mu\nu}{}_{\alpha'\beta'},
\end{equation}
where $\mathcal{P}$ is a bitensor. We then have the linearized metric given by $g_{\mu\nu} = \eta_{\mu\nu} + \sqrt{16\pi G}h_{\mu\nu}$, where
\begin{equation}
    h^{\mu\nu}(\mf x) = \sqrt{4 \pi G}\int \dd V' G_R^{\mu\nu}{}_{\alpha' \beta'}(\mf x,\mf x') T^{\alpha' \beta'}(\mf x'),
\end{equation}
where $T_{\alpha\beta}$ denotes the stress-energy tensor of the source. For the case of a pointlike particle undergoing a trajectory $\mf z_1(t)$  with four-velocity $u_1^\mu(t)$, it reads
\begin{equation}
    T_1^{\mu\nu}(\mf x) = m_1 u_1^\mu(t) u_1^\nu(t) \frac{\delta^{(3)}(\bm x - \bm z_1(t))}{u_1^0(t) \sqrt{-g}},
\end{equation}
so that we obtain
\begin{align}
    \int \dd V' G_R^{\mu\nu}{}_{\alpha'\beta'}(\mf x,\mf x')T_1^{\alpha' \beta'}(\mf x') &= \frac{m_1}{4\pi} \int \dd t' \dd^3 \bm x' \frac{1}{u_1^0(t') |\bm x - \bm x'|} \delta^{(3)}(\bm x' - \bm z_1(t'))\delta(t-t' - |\bm x  - \bm x'|) \Pi^{\mu\nu}{}_{\alpha'\beta'}u_1^{\alpha'}(t') u_1^{\beta'}(t')\nonumber\\
    &= \frac{m_1}{4\pi} \int \dd t'  \frac{1}{u_1^0(t') |\bm x - \bm z_1(t')|} \delta(t-t' - |\bm x  - \bm z_1(t')|) \mathcal{P}^{\mu\nu}{}_{\alpha'\beta'}u_1^{\alpha'}(t') u_1^{\beta'}(t').
\end{align}
Now, let $t_r$ be the retarded time such that $t - t_r - |\bm x - \bm z(t_r)| = 0$, so that
\begin{equation}
    \delta(t-t' - |\bm x  - \bm z_1(t')|) = \frac{\delta(t'-t_r)}{1- \hat{\bm r}_1(t_r)\!\cdot\! \dot{\bm z}_1(t_r)},
\end{equation}
where $\hat{\bm r_1}(t) = (\bm x  - \bm z_1(t))/|\bm x  - \bm z_1(t)|$ and we obtain
\begin{align}
    \int \dd V' G_R^{\mu\nu}{}_{\alpha'\beta'}(\mf x,\mf x')T_1^{\alpha' \beta'}(\mf x') = \frac{m_1}{4\pi}  \frac{\mathcal{P}^{\mu\nu}{}_{\alpha'\beta'}u_1^{\alpha'}(t_r) u_1^{\beta'}(t_r)}{u_1^0(t_r)(1- \hat{\bm r}_1(t_r)\!\cdot\! \dot{\bm z}_1(t_r)) |\bm x - \bm z_1(t_r)|}.
\end{align}
The interaction Hamiltonian of a particle labelled $2$ with the gravitational potential sourced by particle $1$ will then be
\begin{align}
    {H}_{12}(t) = - \frac{1}{2} \sqrt{16\pi G} \int \dd^3 {\bm x}\, h_{\mu\nu}(\mf x) T_2^{\mu\nu}(\mf x) &=- \frac{G m_1 m_2}{|\bm z_2(t) - \bm z_1(t_r)|} \frac{\mathcal{P}_{\mu\nu}{}_{\alpha'\beta'}u_2^{\mu}(t) u_2^{\nu}(t)u_1^{\alpha'}(t_r) u_1^{\beta'}(t_r)}{u_2^0(t)u_1^0(t_{r_{12}})(1- \hat{\bm r}_{12}\!\cdot\! \dot{\bm z}_1(t_{r_{12}}))},\\
    &=- \frac{G m_1 m_2}{|\bm z_2(t) - \bm z_1(t_{r_{12}})|} \frac{2(\eta_{\mu\nu}u_2^{\mu}(t) u_1^{\nu}(t_{r_{12}}))^2-1}{u_2^0(t)u_1^0(t_{r_{12}})(1- \hat{\bm r}_{12}\!\cdot\! \dot{\bm z}_1(t_{r_{12}}))},
\end{align}
where $t_{r_{12}}$ is the solution to $t - t_{r_{12}} = |\bm z_2(t) - \bm z_1(t_{r_{12}})|$ and $\hat{\bm r}_{12} = (\bm z_2(t)  - \bm z_1(t_{r_{12}}))/|\bm z_2(t)  - \bm z_1(t_{r_{12}})|$, and we used $\mathcal{P}_{\mu\nu}{}_{\alpha'\beta'} = \eta_{\mu\alpha'}\eta_{\nu \beta'} + \eta_{\mu \beta'} \eta_{\nu \alpha'} - \eta_{\mu\nu} \eta_{\alpha'\beta'}$. The total interaction between the two particles is then given by
\begin{align}
    H_I(t) = \frac{1}{2}\left(H_{12}(t) + H_{21}(t)\right) =- \frac{G m_1 m_2}{|\bm z_2(t) - \bm z_1(t_{r_{12}})|}& \frac{(\eta_{\mu\nu}u_2^{\mu}(t) u_1^{\nu}(t_{r_{12}}))^2-1/2}{u_2^0(t)u_1^0(t_{r_{12}})(1- \hat{\bm r}_{12}\!\cdot\! \dot{\bm z}_1(t_{r_{12}}))}\\&- \frac{G m_1 m_2}{|\bm z_1(t) - \bm z_2(t_{r_{21}})|} \frac{(\eta_{\mu\nu}u_1^{\mu}(t) u_2^{\nu}(t_{r_{21}}))^2-1/2}{u_1^0(t)u_2^0(t_{r_{21}})(1- \hat{\bm r}_{12}\!\cdot\! \dot{\bm z}_2(t_{r_{21}}))},
\end{align}
$t_{r_{21}}$ is the solution to $t - t_{r_{21}} = |\bm z_2(t) - \bm z_1(t_{r_{21}})|$ and $\hat{\bm r}_{21} = (\bm z_1(t)  - \bm z_2(t_{r_{21}}))/|\bm z_1(t)  - \bm z_2(t_{r_{21}})|$, and we used $\mathcal{P}_{\mu\nu}{}_{\alpha'\beta'} = \eta_{\mu\alpha'}\eta_{\nu \beta'} + \eta_{\mu \beta'} \eta_{\nu \alpha'} - \eta_{\mu\nu} \eta_{\alpha'\beta'}$.
Notice that in the non-relativistic limit, we have $t_{r_{12}}\approx t_{r_{21}} \approx t$, $u_1^0(t) \approx u_2^0(t) \approx 1$ and $\eta_{\mu\nu} u_1^\mu(t)u_2^\nu \approx 1$, allowing one to recover the non-local Newtonian interaction between the particles:
\begin{equation}
    H_I(t)\approx - \frac{G m_1 m_2}{|\bm z_1(t) - \bm z_2(t)|}.
\end{equation}

Also notice that
\begin{align}
    \int \dd t\, H_I(t) &= 2 \pi G\int \dd V \dd V'\left( T^1_{\mu\nu}(\mf x) G_R^{\mu\nu}{}_{\alpha'\beta'}(\mf x, \mf x')T_2^{\alpha' \beta'}(\mf x') +  T^2_{\mu\nu}(\mf x) G_R^{\mu\nu}{}_{\alpha'\beta'}(\mf x, \mf x')T_1^{\alpha' \beta'}(\mf x')\right) \\&= 2 \pi G\int \dd V \dd V'\left( T^1_{\mu\nu}(\mf x) G_R^{\mu\nu}{}_{\alpha'\beta'}(\mf x, \mf x')T_2^{\alpha' \beta'}(\mf x') +  T^2_{\alpha'\beta'}(\mf x') G_R^{\alpha'\beta'}{}_{\mu\nu}(\mf x', \mf x)T_1^{\mu \nu}(\mf x')\right) \\&= 2 \pi G\int \dd V \dd V' T^1_{\mu\nu}(\mf x) \Delta^{\mu\nu}{}_{\alpha'\beta'}(\mf x, \mf x')T_2^{\alpha' \beta'}(\mf x'),
\end{align}
where $\Delta^{\mu\nu\alpha'\beta'}(\mf x, \mf x') = G_R^{\mu\nu\alpha'\beta'}(\mf x, \mf x') + G_A^{\mu\nu\alpha'\beta'}(\mf x,\mf x')$ and we used $G_R^{\alpha'\beta'\mu\nu}(\mf x',\mf x) = G_A^{\mu\nu\alpha'\beta'}(\mf x,\mf x')$.

%\section{Detectors in superposition of trajectories}

%In order to describe detectors in superposition of trajectories, we employ the formalism that is by now well know in the literature~\cite{achimDelocalizedHarvesting,foo}. It consists of associating a quantum degree of freedom to each trajectory, so that the detector interacts with the field at each one of these. That is, in order to consider a detector in superposition of two trajectories we consider a two-dimensional Hilbert space spanned by kets $\{\ket{1},\ket{2}\}$, so that $\ket{1}$ is associated with the trajectory $\mf z_1(\tau_1)$ and $\ket{2}$ is associated with $\mf z_2(\tau_2)$. Then, the interaction of the system with an external gravitational field is given by
%\begin{equation}
    %\hat{\mathcal{H}}_I(\mf x) = - \frac{1}{2} \left({T}^{\mu\nu}_1 \ket{1}\!\!\bra{1}+{T}^{\mu\nu}_2 \ket{2}\!\!\bra{2}\right)\hat{h}_{\mu\nu}(\mf x),
%\end{equation}
%where $T^{\mu\nu}_i$ is the stress energy tensor associated with the $i$th trajectory. If the detector is point-like and the labels $1$ and $2$ are associated to the trajectories $\mf z_1(\tau_1)$ and $\mf z_2(\tau_2)$, we then have
%\begin{equation}
    %T^{\mu\nu}_i = m\frac{\delta^{(3)}(\mf x - \mf z_i(\tau_i))}{\sqrt{-g}}u^\mu_i u^\nu_i.
%\end{equation}

\section{Perturbation theory in the quantum case}\label{app:perturbation}

In this appendix we perform the calculations of the time evolution operator for the quantum treatment of the BMV experiment. For convenience, we write the interaction Hamiltonian that considers the particles coupled to linearized quantum gravity as
\begin{align}\label{eq:Hred}
    \hat{H}_I(t) &= -\sqrt{4\pi G}\sum_{i=1}^2\sum_{p=L,R} \ket{p_i}\!\!\bra{p_i}\,\hat{\phi}_{p_i}(\mf z_{p_i}(t)),
\end{align}
where we defined
\begin{align}
    \hat{\phi}_{p_i}(\mf z_{p_i}(t)) = \int \dd^3 \bm x  m_i\frac{u_{p_i}^{\mu}(t)u_{p_i}^{\nu}(t)}{u_{p_i}^0(t)}\hat{h}_{\mu\nu}(\mf z_{p_i}(t)).
\end{align}
The time evolution operator can be written as
\begin{equation}
    \hat{U}_I = \mathcal{T}\text{exp}\left(-\ii \int \hat{H}_I(t)\right),
\end{equation}
and the Dyson expansion gives
\begin{equation}
    \hat{U}_I  = \openone + \hat{U}^{(1)}_I + \hat{U}^{(2)}_I+\mathcal{O}(3),
\end{equation}
where $\mathcal{O}(3)$ denotes terms of third order in the interaction Hamiltonian. Explicitly,
\begin{align}
    \hat{U}^{(1)}_I = -\ii \int \dd t \hat{H}_I(t), \quad\quad
    \hat{U}_I^{(2)} = - \int \dd t \dd t' \hat{H}_I(t) \hat{H}_I(t') \theta(t-t'),
\end{align}
where $\theta(t-t')$ is the Heaviside step function and implements time ordering. Plugging Eq. \eqref{eq:Hred} for the Hamiltonian, we  obtain
\begin{align}
    \hat{U}_I^{(1)} &= \ii \sqrt{4\pi G} \sum_{p=L,R} \ket{p_1}\!\!\bra{p_1}\,\int \dd t\, \hat{\phi}_{p_1}(\mf z_{p_1}(t))+\ii \sqrt{4\pi G} \ket{p_2}\!\!\bra{p_2}\,\int \dd t \,\hat{\phi}_{p_2}(\mf z_{p_2}(t)),\\
    \hat{U}^{(2)}_I &=- 4\pi G\sum_{p=L,R} \ket{p_1}\!\!\bra{p_1}\int \dd t \dd t'\, \hat{\phi}_{p_1}(\mf z_{p_1}(t))\hat{\phi}_{p_1}(\mf z_{p_1}(t'))\theta(t-t')+\ket{p_2}\!\!\bra{p_2}\int \dd t \dd t' \,\hat{\phi}_{p_2}(\mf z_{p_2}(t))\hat{\phi}_{p_2}(\mf z_{p_2}(t'))\theta(t-t') \nonumber\\&-4\pi G \sum_{p=L,R} \ket{p_1p_2}\!\!\bra{p_1p_2}\int \dd t \dd t'\, \left(\hat{\phi}_{p_1}(\mf z_{p_1}(t))\hat{\phi}_{p_2}(\mf z_{p_2}(t'))\theta(t-t')+\hat{\phi}_{p_2}(\mf z_{p_2}(t))\hat{\phi}_{p_1}(\mf z_{p_1}(t'))\theta(t-t')\right).\\
    \hat{U}^{(2)\dagger}_I &=- 4\pi G\sum_{p=L,R} \ket{p_1}\!\!\bra{p_1}\,\int \dd t \dd t'\, \hat{\phi}_{p_1}(\mf z_{p_1}(t))\hat{\phi}_{p_1}(\mf z_{p_1}(t'))\theta(t'-t)+\ket{p_2}\!\!\bra{p_2}\,\int \dd t \dd t' \,\hat{\phi}_{p_2}(\mf z_{p_2}(t))\hat{\phi}_{p_2}(\mf z_{p_2}(t'))\theta(t'-t) \nonumber\\&-4\pi  G \sum_{p=L,R} \ket{p_1p_2}\!\!\bra{p_1p_2}\int \dd t \dd t'\, \left(\hat{\phi}_{p_2}(\mf z_{p_2}(t))\hat{\phi}_{p_1}(\mf z_{p_1}(t'))\theta(t'-t)+\hat{\phi}_{p_1}(\mf z_{p_1}(t))\hat{\phi}_{p_2}(\mf z_{p_2}(t'))\theta(t'-t)\right),
\end{align}
where we implemented the change of variables $t\leftrightarrow t'$ in the double integrals of $\hat{U}_I^{(2)\dagger}$.

\section{Final states of the particles}

The initial state of the two particles in matrix form in the basis $\{\ket{L_1L_2},\ket{R_1 L_2},\ket{L_1R_2},\ket{R_1R_2}\}$ reads
\begin{equation}
    \hat{\rho} = \frac{1}{4}
    \begin{pmatrix}
        1 & 1 & 1 & 1\\
        1 & 1 & 1 & 1\\
        1 & 1 & 1 & 1\\
        1 & 1 & 1 & 1        
    \end{pmatrix}.
\end{equation}
To leading order in the gravitational field, the updated state of the particles can be written as $\hat{\rho} +  \delta\hat{\rho}_c$, where
\begin{equation}
    \delta\hat{\rho}_c = -\frac{\ii \pi G}{2}
    \begin{pmatrix}
        0 & \Delta_{R_1L_2}-\Delta_{L_1L_2} &\Delta_{L_1R_2}-\Delta_{L_1L_2} & \Delta_{R_1R_2}-\Delta_{L_1L_2}\\
        \Delta_{R_1L_2}-\Delta_{L_1L_2} & 0 & \Delta_{L_1R_2}-\Delta_{R_1L_2} & \Delta_{R_1R_2}-\Delta_{R_1L_2}\\
       \Delta_{L_1L_2}-\Delta_{L_1R_2} & \Delta_{L_1R_2}-\Delta_{R_1L_2} & 0 & \Delta_{L_1R_2}-\Delta_{R_1R_2}\\
        \Delta_{L_1L_2}-\Delta_{R_1R_2} & \Delta_{R_1L_2}-\Delta_{R_1R_2} & \Delta_{L_1R_2}-\Delta_{R_1R_2} & 0      
    \end{pmatrix}.
\end{equation}
In the quantum case, the final state of the particles to leading order can be written as $\hat{\rho} + (\delta\hat{\rho}_c+\delta\hat{\rho}_q+\delta \hat{\rho}_l)$, where
\begin{equation}
    \delta\hat{\rho}_l = \pi G (\mathcal{L}_\textsc{v}-\mathcal{L}_{\textsc{i}})
    \begin{pmatrix}
        0  &  1 & 1 & 2\\
        1  &  0 & 2 & 1\\
        1  &  2 & 0 & 1\\
        2  &  1 & 1 & 0
    \end{pmatrix},
\end{equation}
and
\begin{equation}
    \delta\hat{\rho}_q = \frac{\pi G}{2}(H_{L_1R_2} \!\!+ \!\!H_{R_1L_2} \!\!-\!\! H_{L_1L_2}\!\!-\!\!H_{R_1R_2})
    \begin{pmatrix}
        0 & 0 & 0 & 1\\
        0 & 0 & -1 & 0\\
        0 & -1 & 0 & 0\\
        1 & 0 & 0 & 0
    \end{pmatrix},
\end{equation}
where $H_{p_1p_2}$ denotes the integrated Hadamard function along each pair of paths for particles 1 and 2,
\begin{equation}
    H_{p_1p_2} = \int \dd V \dd V' T^{\mu\nu}_{p_1}H_{\mu\nu\alpha'\beta'}(\mf x,\mf x')T_{p_2}^{\alpha'\beta'}(\mf x').
\end{equation}
while $\mathcal{L}_\textsc{v}$ and $\mathcal{L}_{\textsc{i}}$ are noise terms due to the interaction of each particle with the vacuum of the field. These are explicitly given by
\begin{align}
    \mathcal{L}_{\textsc{v}} &= \int \dd V \dd V' T^{\mu\nu}_{L_i}\langle \hat{h}_{\mu\nu}(\mf x) \hat{h}_{\alpha'\beta'}(\mf x')\rangle_0 T_{L_i}^{\alpha'\beta'}(\mf x')= \int \dd V \dd V' T^{\mu\nu}_{R_i}\langle \hat{h}_{\mu\nu}(\mf x) \hat{h}_{\alpha'\beta'}(\mf x')\rangle_0 T_{R_i}^{\alpha'\beta'}(\mf x'),\\
    \mathcal{L}_{\textsc{i}} &= \int \dd V \dd V' T^{\mu\nu}_{L_i}\langle \hat{h}_{\mu\nu}(\mf x) \hat{h}_{\alpha'\beta'}(\mf x')\rangle_0 T_{R_i}^{\alpha'\beta'}(\mf x'),
\end{align}
where, due to choice of paths, the indices $i$ can be $1$ or $2$, and still yield the same result due to the fact that the paths are related to translations and rotations, which are symmetries of the quantum field theory. Then, the $\mathcal{L}_{\textsc{v}}$ term is a local noise associated to each path, while the $\mathcal{L}_{\textsc{i}}$ term represents an interference term associated with each particle undergoing the superposition of paths. Also notice that due to the fact that the propagator decreases with distance, we have $\mathcal{L}_{\textsc{i}}\leq \mathcal{L}_{\textsc{v}}$, with equality holding only if the paths $1$ and $2$ are identical. Moreover, these vacuum noise terms decay fast with the interaction time, so that they are negligible for long interaction times. Then we can interpret $\delta{\hat{\rho}}_l$ as a local vacuum contribution and $\delta\hat{\rho}_q$ can be seen as the additional correlation contribution due to the quantum nature of the field.

%\end{comment}

\bibliography{references}

\begin{thebibliography}{37}
\bibitem [{Per()}]{Perro}%
  \BibitemOpen
  \href@noop {} {\bibinfo  {journal} {{Notice that in the nomenclature of
  quantum information an interaction between two systems of the
  form~\eqref{eq:Hclass} is called quantum because it establishes a quantum
  channel. However in this manuscript we argue that establishing a quantum
  channel is not a valid proof that the gravitational interaction has a Hilbert
  space structure of any kind or any local quantum degrees of freedom. A proof
  that gravity is quantum requires one to see an experimental phenomenon that
  cannot be explained unless there are local degrees of freedom for the
  gravitational field, as we argue below}}\ }\BibitemShut {NoStop}%
\bibitem [{foo()}]{footNeg}%
  \BibitemOpen
\bibfield  {journal} {  }\href@noop {} {\bibinfo  {journal} {We use negativity
  instead of entanglement entropy to compare this classical calculation with
  the quantum description for the gravitational field below. When the field is
  quantum there will be particle-field entanglement and the state of the
  particles will not be pure. Entanglement entropy is therefore not a valid
  entanglement measure. Negativity is always a faithful entanglement quantifier
  for bipartite two-level systems~\cite{VidalNegativity}}\ }\BibitemShut
  {NoStop}%
\end{thebibliography}

\end{document}